\documentclass[journal]{IEEEtran}

\ifCLASSINFOpdf
\else
\fi

\usepackage[utf8]{inputenc} 
\usepackage{amsmath,graphicx,color}
\usepackage[utf8]{inputenc}
\usepackage{caption}
\usepackage{subcaption}
\usepackage{amsfonts, dsfont}
\usepackage{amssymb}
\usepackage{epstopdf}
\usepackage{siunitx}
\usepackage{booktabs}
\usepackage{cite}
\usepackage{algorithm}
\usepackage[noend]{algpseudocode}
\usepackage{pgfplots} 
\pgfplotsset{compat=newest} 
\pgfplotsset{plot coordinates/math parser=false} 
\newlength\figureheight 
\newlength\figurewidth

\begin{document}

\title{Gravitational Clustering: A Simple, Robust and Adaptive Approach for Distributed Networks}

\author{Patricia~Binder,
        Michael~Muma
        and~Abdelhak~M.~Zoubir}

\maketitle

\begin{abstract}
Distributed signal processing for wireless sensor networks enables that different devices cooperate to solve different signal processing tasks. A crucial first step is to answer the question: who observes what? Recently, several distributed  algorithms have been proposed, which frame the signal/object labelling problem in terms of cluster analysis after extracting source-specific features, however, the number of clusters is assumed to be known. We propose a new method called Gravitational Clustering (GC) to adaptively estimate the time-varying number of clusters based on a set of feature vectors. The key idea is to exploit the physical principle of gravitational force between mass units: streaming-in feature vectors are considered as mass units of fixed position in the feature space, around which mobile mass units are injected at each time instant. The cluster enumeration exploits the fact that the highest attraction on the mobile mass units is exerted by regions with a high density of feature vectors, i.e., gravitational clusters. By sharing estimates among neighboring nodes via a diffusion-adaptation scheme, cooperative and distributed cluster enumeration is achieved. Numerical experiments concerning robustness against outliers, convergence and computational complexity are conducted. The application in a distributed cooperative multi-view camera network illustrates the applicability to real-world problems.
\end{abstract}


\IEEEpeerreviewmaketitle

\section{Introduction}\label{sec:intro}

\IEEEPARstart{}{}

\IEEEPARstart{A}{}recent and emerging research direction in distributed signal processing for wireless sensor networks (WSNs) is that of enabling cooperation among multiple heterogeneous devices dedicated to solve different signal processing tasks \cite{plata2017heterogeneous,HASSANI201568,TONG2016149,Chen201635, SZURLEY201544}.  A crucial first step towards this so-called multiple devices multiple tasks (MDMT) paradigm is to answer the question: who observes what? \cite{Teklehaymanot2015, Chouvardas2015,bahari2016distributed, BERTRAND20121679}. For example,  to arrive at a node-specific speech signal enhancement \cite{bertrand2011distributed,hassani2016multi,hassani2017multi}, all relevant speech sources must be uniquely labelled throughout the wireless acoustic sensor network. Similarly, distributed node-specific image/video enhancement requires the common labelling of all objects within a camera network \cite{Teklehaymanot2015}. 

To illustrate the challenging requirements for such labelling methods, consider for example a video enhancement setup, where multiple users film a nonstationary scene from different angles using their camera equipped portable devices. Each user has its own dedicated signal processing task, e.g. enhancing a specific object of interest. No prior information, such as positions of devices, registration of views or number of objects in the scene, is available, streaming-in data must be processed sequentially and little is known about the distribution of the data. Further, a central computing unit (fusion center) is not available and communication (range, bandwidth) and computation capabilities (memory, computing power), as well as battery power may be limited.

Recently, several  distributed  algorithms have  been  proposed, which frame the labelling problem in terms of cluster analysis after extracting source-specific features \cite{Chouvardas2015,Binder2015,Teklehaymanot2015, Binder2016,bahari2016distributed}.  Various methods have been proposed for distributed data clustering, e.g., \cite{HaiSurvey,Nowak,DKM,7065284,6232902, tubiblio74210,Binder2016, 6375088,shen2014distributed,s120100092,7425183, 7475096,gu2008distributed, Pastor}. However, a significant drawback of common clustering algorithms is that the number of clusters has to be known a priori. In real scenarios, this information is not always available\cite{issac2014case} or the number of clusters might be chosen improperly. Also, in a sensor network the number of clusters may change over time in a non-stationary scenario.

To the best of our knowledge, distributed cluster enumeration has only been addressed in \cite{Tekle_distrXMeans}, which serves here as a benchmark algorithm. For the single-node case, the question of inferring the number of clusters from the observations has been intensively studied \cite{howManyClusters, 1374239, BIOM784, XMeans, PGMeans, Milligan1985, Boutsinas2006, refId0, FRIGUI19961223, Fang2012468, Kolesnikov2015941,  biernacki1997using, Qian2009, 4515866, MeanShift, MeanShift2, Dudoit2002, Kothari1999405, 5383365, 1532874, Nakamura19981265, Zhao2008, YU2014101, Herbin20011557}.
However, most of these approaches are of high computational complexity, need to make prior assumptions on the data distribution or do not allow for an adaptive processing without the need to re-run the entire algorithm. Therefore, these methods are not suitable for the above-described object labelling task in MDMT networks. 

The aim of this research is to adaptively estimate the time-varying number of clusters based on a set of streaming-in feature vectors. The proposed method is designed to be
\begin{enumerate}
\item {\it adaptive} - to a changing number of objects/sources,
\item {\it robust} - against outliers in the feature vectors, or in general against unknown non-spherical and possibly heavy tailed distributions of the estimated features,
\item {\it distributed} -  so as to operate in a decentralized WSN, e.g., based on the diffusion-principle \cite{sayed2014, Binder2016},
\item {\it sequential} - so that the estimate of the number of clusters is continuously updated for streaming-in data without the need to re-run the entire algorithm,
\item {\it computationally simple} - in order to be applicable in a real WSN.
\end{enumerate}

\textbf{Original Contributions:} A robust gravitational clustering algorithm is proposed which works for single-node and cooperative in-network clustering. 
The key idea is to exploit the physical principle of gravitational force between mass units. In this work, streaming-in feature vectors are considered as mass units of a fixed position in the feature space, around which mobile mass units are injected at each time instant. The cluster enumeration exploits the fact that the highest attraction on the mobile mass units is exerted by regions with a high density of feature vectors, i.e., gravitational clusters. The masses of mobile units are combined when they are in a close vicinity of each other and a threshold on the combined mass serves as detector for a cluster. In this way, the time-varying number of clusters can be determined. By sharing estimates among neighboring nodes via a diffusion-adaptation scheme, cooperative and distributed cluster enumeration is achieved. An extensive simulation-based performance analysis is provided that investigates the clustering performance for single-node and multi-node cluster enumeration. Herein, aspects such as robustness against outliers, computational cost and convergence are investigated. The applicability of the gravitational clustering algorithm is illustrated for a use-case of labelling moving objects in a synthetic 3-D multi camera network. 

{\bf Related Work:} The idea to cluster data based on the law of gravity was first proposed by Wright in \cite{WRIGHTGravitationalClustering} where clustering is performed by moving and merging the data points based on gravitational force until one final cluster remains. This approach has been extended in some works, e.g., \cite{gomez2003new,Sanchez2014498}, which consider multiple clusters. An overview of existing methods which exploit the gravitational principle is provided in \cite{Sanchez2014498}. In these methods, a decay term prevents that all samples conflate into one big cluster, or a threshold is set that determines up to which distances clusters should stay separated and which clusters may merge. This requires prior knowledge about the data, e.g., the minimum distance between the clusters or the distance of the data points from their corresponding cluster centroid, in order to assure adequate performance. This kind of information is not always available. Another drawback is that one cannot draw inferences from the resulting clusters about the actual positions of the cluster centroids since the data points (and therefore the cluster centroids) change their positions because of their mutual attraction. Further, such a procedure makes it difficult to adapt to changes in the scenario without the need to re-run the algorithm.

\textbf{Notation:} The following notation is used: vectors are denoted by bold small letters $\mathbf{a}$ and matrices by bold capital letters $\mathbf{A}$. All vectors are defined as column vectors.
Sets are denoted by calligraphic letters $\mathcal{A}$ with $|\cdot|$ indicating the cardinality of a set, the notation $\mathcal{A} \setminus i$ describes the resulting set after excluding element $i$ from $\mathcal{A}$ while $\Vert\cdot\Vert$ denotes the Euclidean norm of a vector.
The superscript $^\top$ denotes the transpose operator and $\mathbf{I}_M$ stands for an $M \times M$ identity matrix.\\
\textbf{Organization:} Section II provides the problem formulation and data model. Section III is dedicated to the proposal of our gravitational clustering algorithm, while Section IV provides an extensive Monte-Carlo simulation study. Section V concludes the paper and provides future research directions.

\section{Problem Formulation, Signal Model and Aims} \label{sec:problem}

We consider a network of $J$ nodes whose topology is described by a graph with nodes indexed by $j \in {1,...,J}$. The neighborhood of node $j$, denoted as $\mathcal{B}_j$, is the set of nodes, including $j$, that node $j$ exchanges information with, and $|\mathcal{B}_j|$ denotes its associated cardinality. Each observation is assumed to belong to a certain cluster $\mathcal{C}_k$ with $k \in {1,...,K}$ denoting the label of the given cluster. The total number of clusters $K$ is assumed to be unknown and might change over time. Each cluster is described by a set of application-dependent descriptive statistics (features). 

The feature estimation process is an application-specific research area of its own (see, e.g., \cite{Chouvardas2015, Teklehaymanot2015}) and is not the focus of this article, where we seek for a generic adaptive and robust cluster enumeration method. It is assumed that the features have already been extracted and the uncertainty within each cluster $k$ can be modeled by a probability distribution, e.g., the Gaussian. Further, we account for gross estimation errors in the feature extraction process that we consider as outliers, thus arriving at the following observation model for feature vectors of node $j$ at time instant $t=1,...,N$:
\begin{equation}
\label{eq:regmod}
\mathbf{d}_{kj}(t)= \boldsymbol{w}_k(t) +\boldsymbol{n}_{kj}(t).
\end{equation}
Here, $\boldsymbol{w}_k(t) $ denotes the class centroid, $\boldsymbol{n}_{kj}(t)$ represents a stochastic, clusters-specific uncertainty term of unspecified distribution with associated covariance matrix $\boldsymbol{\Sigma}_{jk}$, and $\boldsymbol{d}_{kj}(t),\boldsymbol{w}_k(t),\boldsymbol{n}_{kj}(t) \in \mathds{R}^{q \times 1}$.
For reasons of visual clarity, we drop the index $k$ in the feature vectors and refer to them as $\boldsymbol{d}_j(t)$. \\

The aim of this research is to estimate the time-varying number of clusters $K(t)$ and class centroids  $\boldsymbol{w}_k(t) $ based on a set of streaming-in feature vectors $\mathbf{d}_j(t)$. The proposed method should be {\it adaptive, robust, distributed, sequential and computationally simple}, as defined in Section~\ref{sec:intro}.

%
%

\section{Description of the Proposed Gravitational Clustering (GC) Algorithm} \label{sec:AlgDes}

Gravitational Clustering (GC) is based on Isaac Newton's law of universal gravitation which relates the force $\boldsymbol{f}$ between two mass units with masses $m_1$ and $m_2$ and distance $\Vert \boldsymbol{r}_{12}\Vert$ by
\begin{align*}
\label{eq:coulomb}
  \boldsymbol{f}_{12}=-\boldsymbol{f}_{21} 
  &= g \cdot m_1 \cdot m_2 \cdot \frac{\boldsymbol{e}_{12}}{\| \boldsymbol{r}_{12}\|^2},
\end{align*} 
where $g=6.67408 \times 10^{11} \si{\metre^3\kilogram^{-1}\second^{-2}}$ is the gravitational constant and $\boldsymbol{e}_{12}$ denotes the unit vector which points from body $1$ to body $2$.  
GC exploits Newton's law by modeling the feature space as a physical space, where $g$ becomes a tuning parameter. 
In contrast to the physical model and also in contrast to \cite{WRIGHTGravitationalClustering,gomez2003new,Sanchez2014498}, two different types of mass units are introduced in this work: on the one hand, the feature vectors are modeled as mass units of fixed position ({\it fixed mass units}).
On the other hand, artificially generated mass units ({\it mobile mass units}) are injected into feature space. Mobile mass units are attracted by fixed mass units, but do not interact with each other.

GC exploits the fact that the highest attraction on the mobile mass units is exerted by regions with a high density of feature vectors, i.e., gravitational clusters, to which mobile mass units gravitate and where they finally gather, governed by
\begin{equation} \label{eq:distance}
\boldsymbol{x}(t)=\int_{t-1}^{t} \boldsymbol{v}(\tau) \text{d}\tau +\boldsymbol{x}(t-1).
\end{equation}
Here, $\boldsymbol{x}(t)$ is the new position of the mobile mass unit at time instance $t$ to which it traveled in time $\Delta t=1$ with velocity $\boldsymbol{v}(t)$ and current position $\boldsymbol{x}(t-1)$.\\ 

The formula for the velocity is given by
\begin{equation} \label{eq:velo}
\boldsymbol{v}(t)=\int_{t-1}^{t} \boldsymbol{a}(\tau) \text{d}\tau +\boldsymbol{v}(t-1),
\end{equation}
with $\boldsymbol{a}(t)$ denoting the acceleration and $\boldsymbol{v}(t-1)$ the current velocity. Using the relation
\begin{equation} \label{eq:Fma}
\boldsymbol{f}(t)=m \cdot \boldsymbol{a}(t)  ,
\end{equation}
solving for $\boldsymbol{a}$, and inserting the result in Eq. \eqref{eq:velo} yields
\begin{equation} \label{eq:velo2}
\boldsymbol{v}(t)=\frac{1}{m} \int_{t-1}^{t} \boldsymbol{f}(\tau)\text{d}\tau +\boldsymbol{v}(t-1).
\end{equation}

Equations \eqref{eq:distance}-\eqref{eq:velo2} provide the basis for the algorithm proposed in this article. In the following subsections, an approach is presented which introduces the clustering procedure for the single-node case as well as the extension to distributed processing for the use in WSNs. Single-node GC is a stand-alone method, i.e., it can be applied to any single device clustering task and does not require a WSN.

\subsection{Single Node Approach for Gravitational Clustering}

We begin by describing the single node approach. For every fixed mass unit with position $\mathbf{d}_j$, which is available at time $t$, a mobile mass unit $\boldsymbol{u}_i$ is emitted in a certain distance $r_x$ around the fixed mass unit in space. The mobile mass unit is characterized by its position $\boldsymbol{x}_i\in \mathds{R}^q$ and its mass $m_{i}$:
\begin{equation*} \label{eq:mmu}
\boldsymbol{u}_i(t)=\left( \boldsymbol{x}_i(t)^\top,\ m_{i}(t) \right)^\top, \quad i \in \mathcal{U}(t).
\end{equation*}
Here, $\mathcal{U}(t)$ denotes the set of indices of all mobile mass units in a feature space at time $t$ with its cardinality  $| \mathcal{U}(t) |$. All fixed mass units are assumed to have equal mass, i.e., $m_d=1$, and the mobile mass units $\boldsymbol{u}_i(0)$ have the initial mass $m_{i}(0)=1$. The distance $r_x$ is a design parameter which must fulfill $r_x>0$ so that the mobile entity is not directly absorbed by the feature vector while being small enough such that the feature vector has an impact on it.\\

The force acting on a mobile mass unit is the superposition of all single forces emanating from each of the fixed mass units $\mathbf{d}_j(t)$. The formula for the total vectorial force acting on a single mobile mass unit $\boldsymbol{u}_i(t)$ is then given by
\begin{equation} \label{eq:totalForce}
\boldsymbol{f}_{\text{grav},i}(t)=\sum_{n=1}^t g \cdot m_{i}(t) \cdot m_d \cdot  \frac{\boldsymbol{d}_j(n)-\boldsymbol{x}_i(t)}{\| \boldsymbol{d}_j(n)-\boldsymbol{x}_i(t)\|^p},
\end{equation}
with $p=3$ according to the physical model. Since we are not restricted to the physical model, $p$ is treated as a further design parameter. \\

To allow a discrete-time representation, we adapt Eqs. (\ref{eq:distance}) and (\ref{eq:velo2}) such that the new position of each attracted entity $\boldsymbol{u}_i(t)$ is calculated based on the following equations:
\begin{equation} \label{eq:distanceAlg}
\boldsymbol{x}_i(t)=\boldsymbol{v}_i(t)\cdot \Delta t +\boldsymbol{x}_i(t-1),
\end{equation}
given the velocity
\begin{equation} \label{eq:veloAlg}
\boldsymbol{v}_i(t)=\frac{\boldsymbol{f}_{\text{grav},i}(t)}{m_{i}(t)} \cdot \Delta t +\boldsymbol{v}_i(t-1)
\end{equation}
with $\boldsymbol{v}_i(0):=0$ for initialization.\\

With decreasing distance to the feature vectors, the force acting on the mass units grows and strives to infinity as the distance goes to zero. As a consequence, and with $\Delta t$ not being chosen infinitesimally small, the mobile mass units are accelerated so strongly that they would ``shoot'' past the data clusters if $\boldsymbol{f}_{\text{grav},i}(t)$ is not limited. For this purpose a viscous damping force $\boldsymbol{f}_{\text{damp},i}(t)$ is introduced which is antagonistic to $\boldsymbol{f}_{\text{grav},i}(t)$, i.e.,
\begin{equation}
\label{eq:Fviscous}
\boldsymbol{f}_{\text{damp},i}(t)=-k_{\text{damp}} \cdot \boldsymbol{v}_i(t-1).
\end{equation}

Here, $k_{\text{damp}}$ is a damping parameter which is chosen such that $0<k_{\text{damp}}< 1$ to ensure the resulting total velocity $\boldsymbol{v}_i(t)$ of the moving mass units is positive.\\ 

Combining Eqs. (\ref{eq:veloAlg}) and (\ref{eq:Fviscous}) yields
\begin{equation} 
\label{eq:veloAlgFinal}
\boldsymbol{v}_i(t)=\frac{\boldsymbol{f}_{\text{grav},i}(t)+ \boldsymbol{f}_{\text{damp},i}(t)}{m_{i}(t)} \cdot \Delta t +\boldsymbol{v}_i(t-1).
\end{equation}
In order to reduce computational cost, for every time step, the algorithm combines any two mobile mass units to a single unit if their distance is equal or smaller than a small constant $\epsilon_r$, which should be chosen significantly smaller than the minimum expected distances between the clusters.

For this purpose, we calculate a $|\mathcal{U}(t)| \times |\mathcal{U}(t)|$ distance matrix $\mathbf{R}(t)$ with elements $r_{m,n}(t), \ m,n=1,...,|\mathcal{U}(t)|$ representing the distance between the mobile mass units:
\begin{equation} \label{eq:distMat}
r_{m,n}(t)=\Vert \boldsymbol{x}_m(t) - \boldsymbol{x}_{n}(t) \Vert , \quad \forall \text{ }m,\text{ } n \in \mathcal{U}(t).
\end{equation}
If two mobile mass units $\boldsymbol{u}_i(t)$ and $\boldsymbol{u}_{i'}(t)$ lie within a distance $\epsilon_r$, i.e. $r_{i,i'}(t) \leq \epsilon_r$, the masses of the combined mobile mass units are summed up and the position of the mobile mass unit, which has the smallest mean distance to all other mobile mass units
\begin{equation} \label{eq:meandist}
r_{\text{mean},i}(t)=\frac{1}{|\mathcal{U}(t)|}  \sum_{\forall j \in \mathcal{U}(t)} r_{ij}(t)	
\end{equation}
is retained, since it is more likely to be combined with other mass units in the next iteration. The other mobile mass unit is removed from the feature space. This procedure is repeated for all mobile mass units that lie within an $\epsilon_r$-distance. After the combination process is finished, all remaining mass units obtain a new indexing such that $\mathcal{U}(t+1) \in \{ 1,2,...,|\mathcal{U}(t+1)| \}$.  Details are given in Algorithm~\ref{combPart}.\\

The mobile mass units continue to move towards the fixed mass units which have the highest attraction and finally remain at the positions where a balance of forces is attained, and hence $\boldsymbol{v}_i(t)\approx 0$. Note that $\boldsymbol{v}_i(t) \rightarrow 0 \text{ for }\Delta t \rightarrow 0$ for the state of balance of forces.. \\

In order to further reduce computational cost, $\boldsymbol{f}_{\text{grav},i}(t)$ can be set to $0$ if $\| \boldsymbol{d}_j(t)-\boldsymbol{x}_i(t)\|>d_{\text{max}}$ with $d_{\text{max}}$ being an arbitrarily large constant.\\

\begin{figure}%
\centering
\begin{subfigure}[b]{.7\linewidth}
 \includegraphics[width=\textwidth]{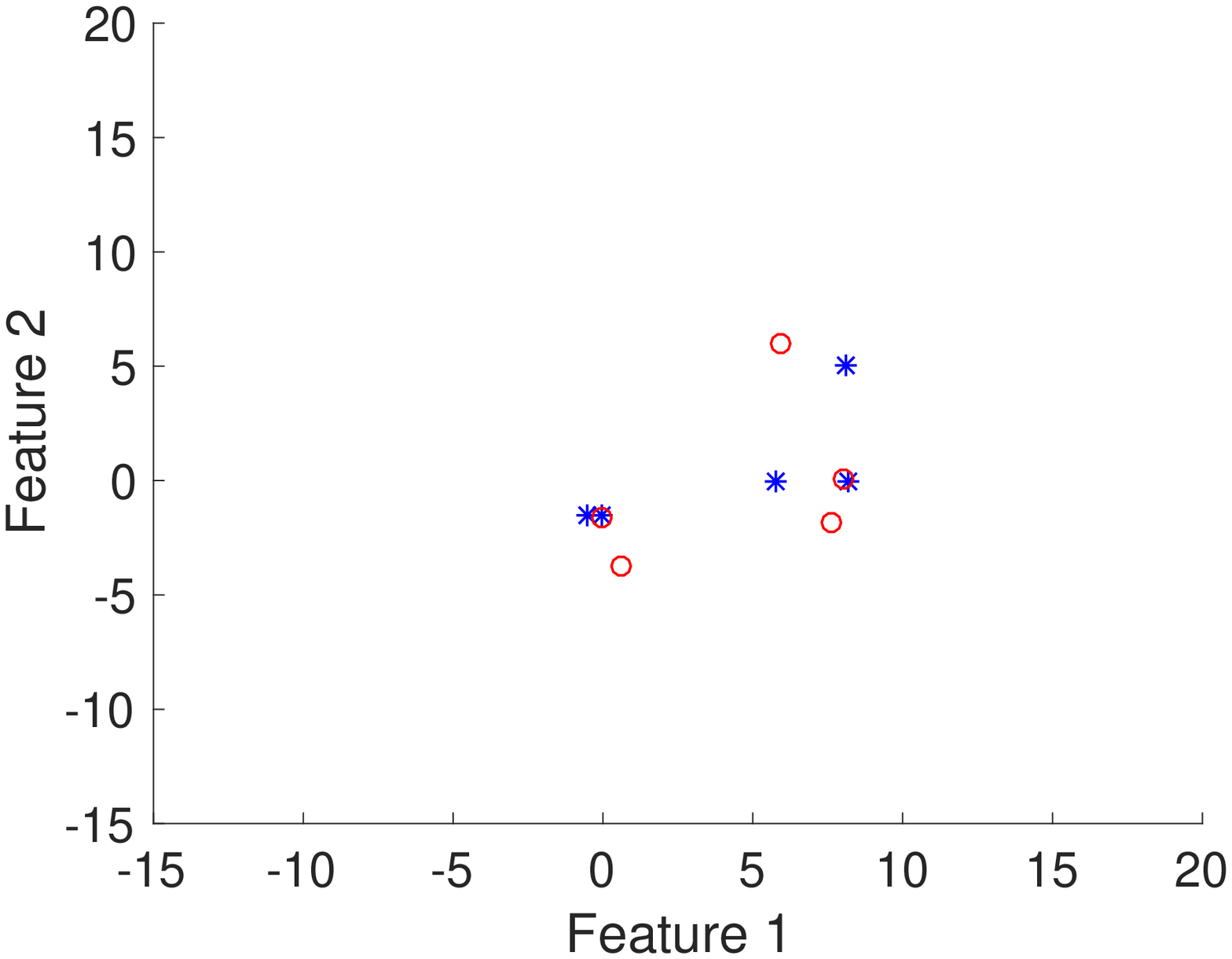}
   \caption{$t=t_0$, $\hat{K}(t)=0$}
\end{subfigure}
\hfill
\begin{subfigure}[b]{.7\linewidth}
  \includegraphics[width=\textwidth]{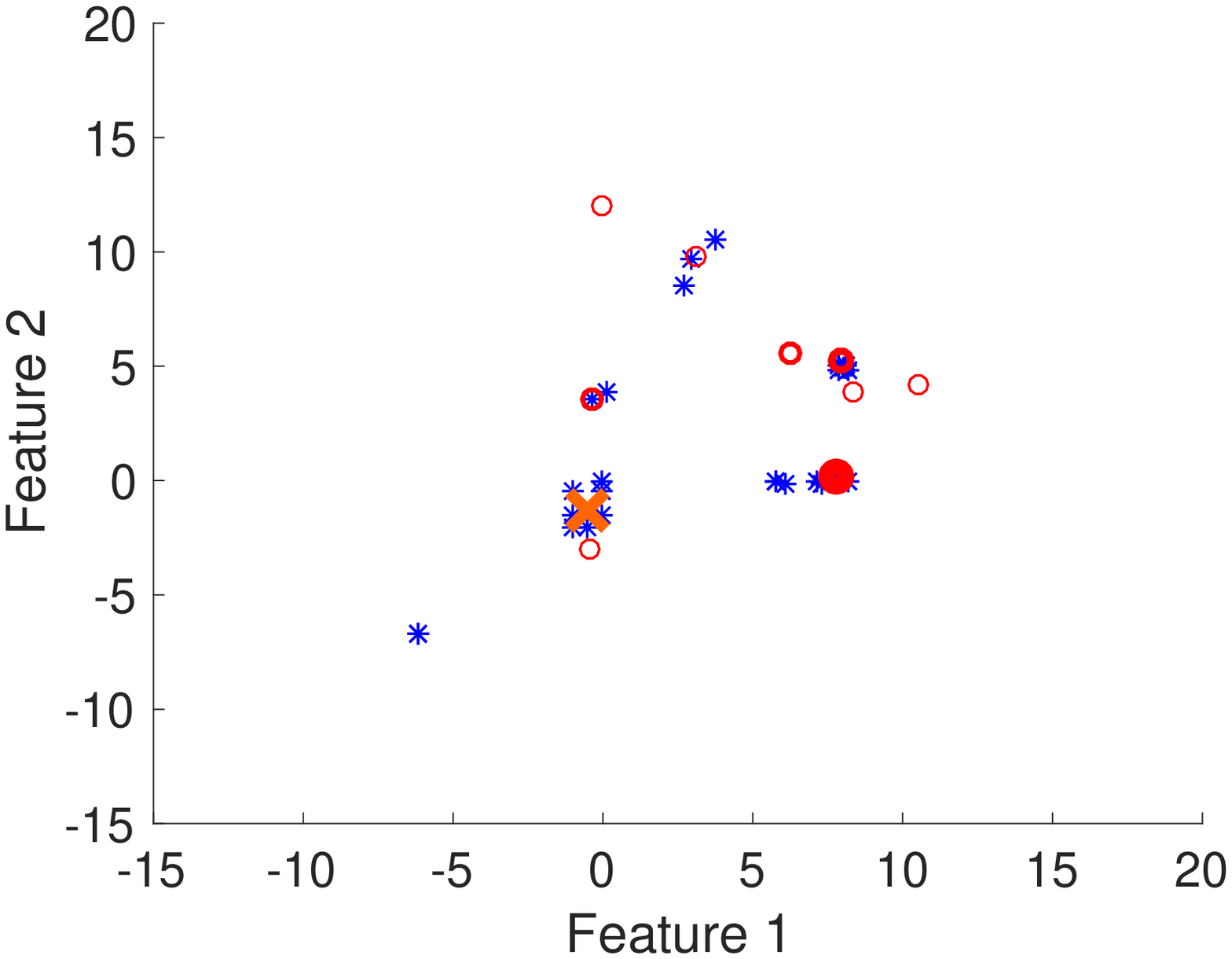}
  \caption{$t=t_1>t_0$, $\hat{K}(t)=1$}
\end{subfigure}

\begin{subfigure}[b]{.7\linewidth}
 \includegraphics[width=\textwidth]{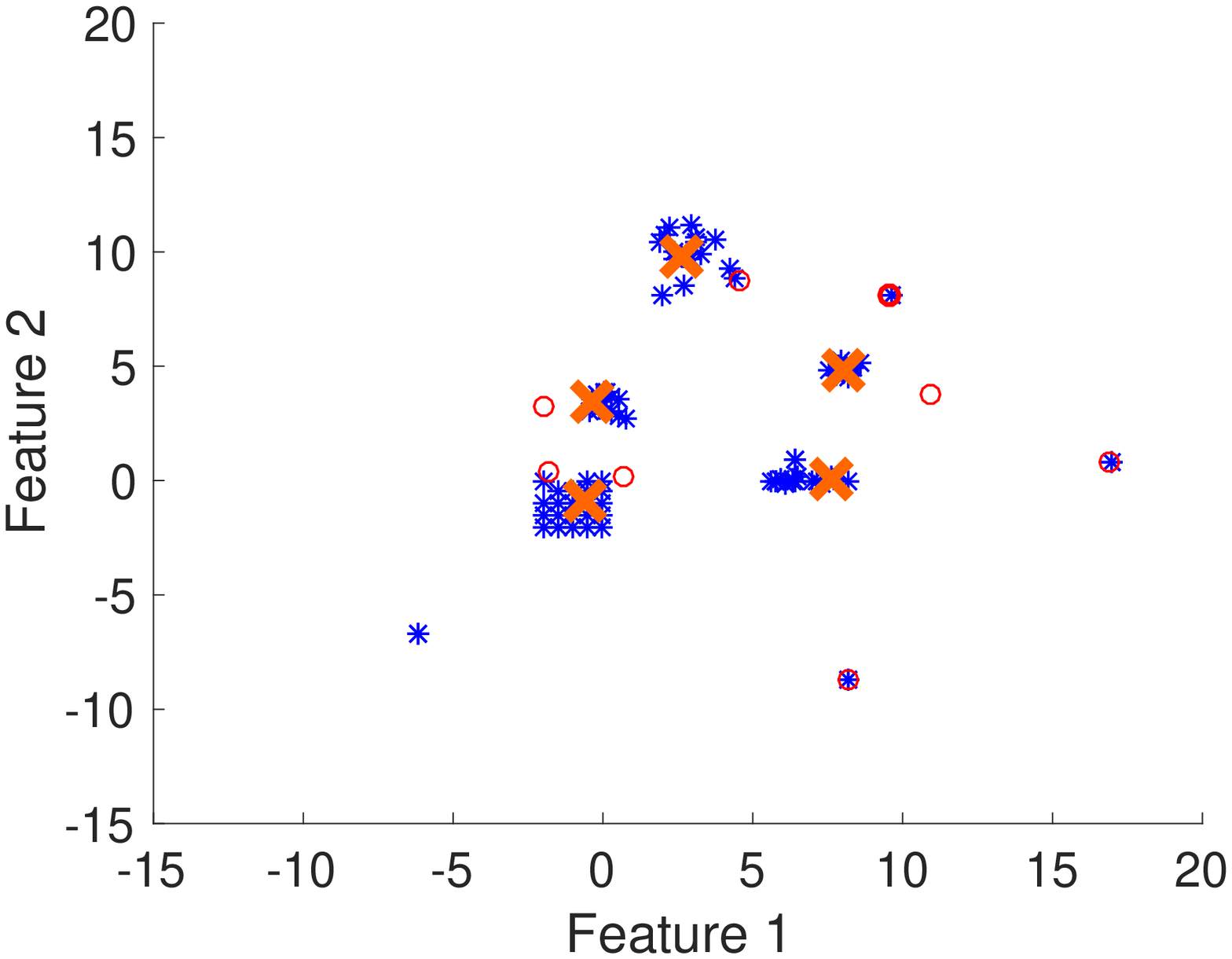}
  \caption{$t=t_2>t_1$, $\hat{K}(t)=5$}
\end{subfigure}
\hfill
\begin{subfigure}[b]{.7\linewidth}
  \includegraphics[width=\textwidth]{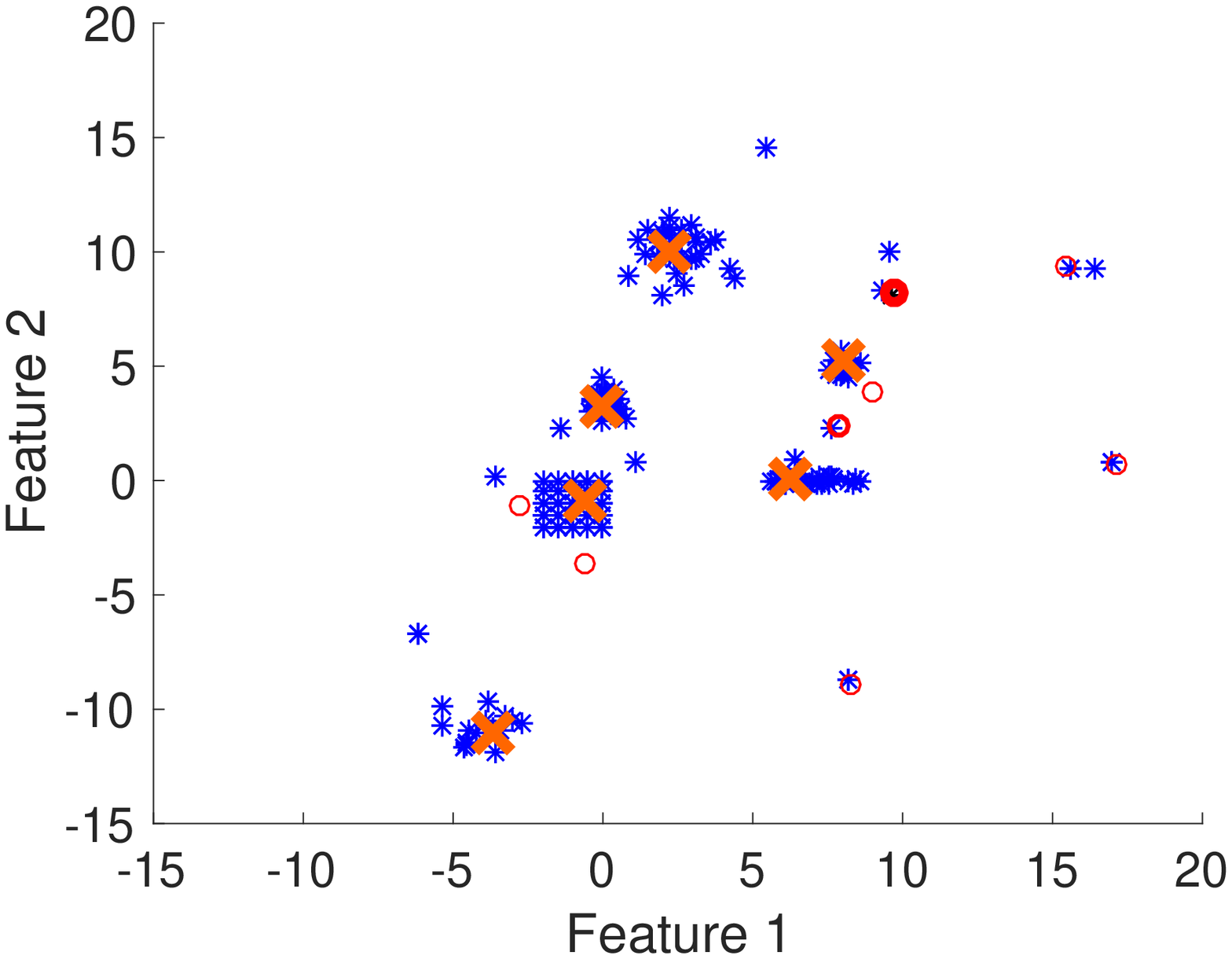}
  \caption{$t=t_3>t_2$, $\hat{K}(t)=6$}
\end{subfigure}
\caption{Exemplary procedure of the GC at different time instants with $p=2$ and 5 $\%$ outliers. The red circles denote the mobile mass units and the blue stars represent the fixed mass units. The thickness of the red circles represents the mass of the units. Once a cluster is found, the respective mobile mass unit is indicated by a large red colored cross. }
\label{Fig:GravClust_StreamingIn}

\end{figure}

\textit{Determination of the Number of Clusters:} \\
In non-stationary scenarios, the number of clusters $K$ depends on the time instant $t$. For every time instant (or for every predefined time interval), the algorithm determines if a mobile mass unit indicates a cluster by checking whether its mass exceeds a threshold $m_{\text{min}}$. This is done for all mass entities such that 
\begin{equation} \label{eq:setmassthresh}
\mathcal{K}(t)=  \{i \mid m_{i}(t)\geq m_{\text{min}}\}, \quad \forall i \in \mathcal{U}(t)
\end{equation}
and an estimate of $K(t)$ is obtained by
\begin{equation} \label{eq:massthresh}
\hat{K}(t)= |\mathcal{K}(t) |.
\end{equation}
Choosing $m_{\text{min}}>1$ prevents that single, ``stuck'' units are misinterpreted as a cluster. As a consequence, single outliers have no influence on the clustering performance. 
 The position of the combined mass units, which indicate a cluster according to Eq. (\ref{eq:massthresh}), provide an estimate of the cluster centroids as defined in Eq. (\ref{eq:regmod}):
 \begin{equation} \label{eq:estcent}
\hat{\boldsymbol{w}}_k(t)=\boldsymbol{x}_k(t), \quad \forall \ k \in \mathcal{K}(t).
\end{equation}
 
The complete gravitational clustering procedure is exemplarily presented in Fig.~\ref{Fig:GravClust_StreamingIn} for different time instants with $p=2$, $\epsilon_r=1$, $r_x=2$, $m_{\text{min}}=7$ and 5 $\%$ outliers, where, at first, data samples from $5$ clusters of different shapes are streaming in one at a time. Starting at $t=t_0$, for each feature vector (represented by the blue stars) available at that time, an associated mobile mass unit (denoted by the red circles) drawn at random from a Gaussian distribution with $\mathcal{N}(\mathbf{d}_n,r_x\mathbf{I}_q)$ (see subfigure (a)) is emitted. With increasing time, the feature vectors begin to form clusters and the mobile mass units move towards the feature vectors. When they gather in the cluster centers, they eventually fuse with other mobile mass units in their direct environment if their distance is less or equal to $\epsilon_r$. In Fig.~\ref{Fig:GravClust_StreamingIn}, this is indicated by the thickness of the red circles which is proportional to the mass of the units. Once a cluster is found according to Eq. (\ref{eq:setmassthresh}), the respective mobile mass unit is represented by a big red colored cross (see subfigure (b)). At $t=t_2$, all $5$ clusters have formed and have been detected by the GC algorithm, i.e. $\hat{K}(t_2)=5$ (see subfigure (c)). At time instant $t$ with $t_2<t<t_3$, a new cluster with centroid $\mathbf{w}_5=(-4, -11)^\top$ is generated additionally to the already existing ones so that from now on data samples from $6$ clusters are streaming in. After a few further time steps, GC adapts to the changing scenario and provides the correct estimate of the number of clusters in the scene (see subfigure (d)). This ability makes it useful for dynamic scenarios where new objects enter the scene, such as in multi-view camera networks. 
This example also illustrates that the algorithm is robust to a certain amount of outliers which is further evaluated in Subsections \ref{sec:robOut} and \ref{sec:SimRob}.\\
\\
A summary of the GC algorithm is provided in Algorithm \ref{GC1}.
\\
\begin{algorithm}
\caption{Combine mobile mass units}\label{combPart}
\begin{algorithmic}[1]
\State calculate distance matrix $\mathbf{R}(t)$
\State $\mathcal{R}(t)=\{ r_{m,n}(t)\mid  r_{m,n}(t) \leq \epsilon_r \},\forall \text{ }m, n \in \mathcal{U}(t),\ m \neq n$        
     \While{$\mathcal{R}(t) \neq \{\}$}
       \State select $r_{\text{min}}=\min (\mathcal{R}(t)):= r_{i,i'}(t)$
       \State calculate Eq. (\ref{eq:meandist}) for $\boldsymbol{u}_i(t)$ and $\boldsymbol{u}_{i'}(t)$
       \If{$r_{\text{mean},i}(t)<r_{\text{mean},i'}(t)$}
        \State $\boldsymbol{u}_i(t+1)=\left( \boldsymbol{x}_i(t)^\top,\ m_{i}(t)+ m_{i'}(t)\right)^\top$
       \Else 
         \State $\boldsymbol{u}_i(t+1)=\left( \boldsymbol{x}_{i'}(t)^\top, \ m_{i}(t)+ m_{i'}(t) \right)^\top$
       \EndIf
       \State $\mathcal{R}(t)\leftarrow\mathcal{R}(t) \setminus r_{i,i'}(t)$

       \State $\mathcal{U}(t+1)\leftarrow\mathcal{U}(t) \setminus i'$
            \EndWhile
\State renew indexing: $\mathcal{U}(t+1) \in \{ 1,2,...,|\mathcal{U}(t+1)| \}$
\end{algorithmic}
\end{algorithm}

\subsubsection{Robustness Against Outliers} \label{sec:robOut}
\begin{figure}
    \centering
\scalebox{0.45}{\input{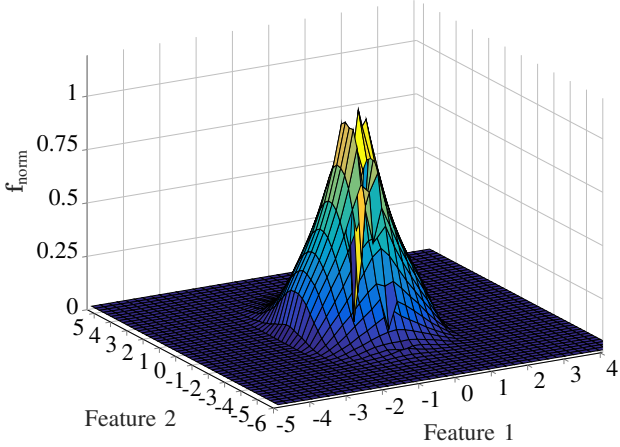}}
    \caption{Normalized force of an outlier for different distances to the cluster centroid in both directions of the two-dimensional feature space.}
    \label{Fig:impOut}
\end{figure}

The reason for the intrinsic robustness against outliers of the GC is visualized in Fig.~\ref{Fig:impOut}, which displays the normalized force that an outlier executes on a cluster. In this example, a single cluster with $500$ feature vectors has its center in $\boldsymbol{w}_1=(0, 0)^\top$ and
covariance matrix $ \boldsymbol{\Sigma}_1=(0.2, 0.4)^\top \mathbf{I}_2$. A single feature vector is moved away stepwise from a cluster of features to evaluate its influence on the elements in the cluster. The normalized force acting on the outlier is evaluated for each position in two-dimensional feature space is shown in Fig.~\ref{Fig:impOut}. While being in the center of the cluster, the outlier experiences equilibrium of forces. The superposition of forces acting on the outlier increases with growing distance from the centroid and reaches its maximum when leaving the bulk of the feature vectors. After that, the influence that the outlier has on the field of gravitational forces decreases continuously with increasing distance to the cluster. Such a bounded influence of outliers on the (cluster) estimates is a desired property of robust methods \cite{RobEst} and is inherent to GC.

\begin{algorithm}
\caption{Gravitational Clustering (GC)}\label{GC1}
\begin{algorithmic}[1]

\For {$t=1,..,N$}
   \State create new mass unit $\boldsymbol{u}_i(t)$ and place around $\boldsymbol{d}_j(t)$
        
        \ForAll {$\boldsymbol{x}_i(t) $}
          \State calculate the total force acting on $\boldsymbol{u}_i(t)$ via Eq. (\ref{eq:totalForce})
          \State assign  $\boldsymbol{u}_i(t)$ its velocity via Eq. (\ref{eq:veloAlgFinal})
          \State determine the new position of $\boldsymbol{u}_i(t)$ via Eq. (\ref{eq:distanceAlg})
        \EndFor
    \State combine mobile mass units according to Algorithm \ref{combPart}
    \State determine $\hat{K}(t)$ via Eqs. (\ref{eq:setmassthresh}) and (\ref{eq:massthresh})
    \State estimate cluster centroids via Eq. (\ref{eq:estcent}) 
\EndFor

\end{algorithmic}
\end{algorithm}

\subsection{Distributed Gravitational Clustering (D-GC)} \label{sec:DGC}
\begin{figure}
\centering
\def\svgwidth{.47\textwidth}
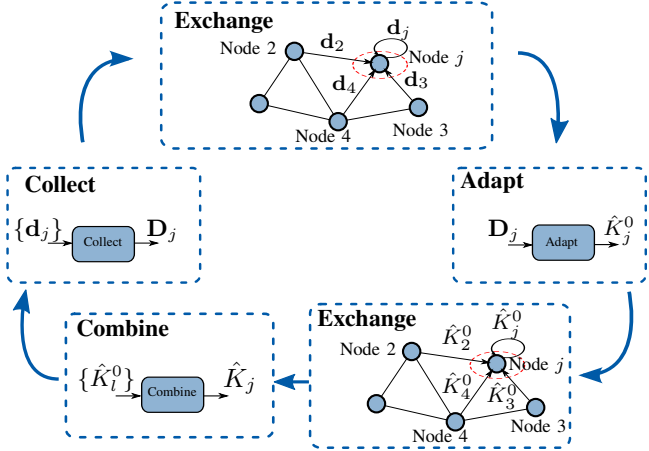
\caption{An overview of the distributed diffusion-based clustering procedure.}
\label{Fig:ATCDiff}
\end{figure}

Consider a WSN with $J$ nodes as defined in Section \ref{sec:problem}.
The task of the WSN is to solve a cluster enumeration problem jointly by communicating the results within the sensor neighborhood, $\mathcal{B}_j$, $j=1,...,J$. The proposed procedure for a cooperative network clustering procedure extends GC by the distributed adaptive communication scheme provided in Fig.~\ref{Fig:ATCDiff} (time indices are left out for reasons of visual clarity). Since the D-GC is based on the diffusion adaptation scheme \cite{sayed2014}, it does not require a fusion center. Instead information diffuses via local neighborhood communication throughout the network. As a result, the adaptation and learning performance is increased compared to non-cooperative sensor networks \cite{diffadapt}.
 
 Having observed $t$ feature vectors $\mathbf{d}_{j}$ at time instant $t$, each node $j$ collects them in a $q \times t$ dimensional matrix $\mathbf{D}_{j}(t)$, where each column consists of an observed feature vector. Optionally, $\mathbf{d}_{j}(t)$ is exchanged within $\mathcal{B}_j$ before adapting, i.e., the number of clusters is determined based on $\mathbf{D}_{j}(t)$ which additionally contains all available $\mathbf{d}_{l}, l\in \mathcal{B}_j$. In this case, the gravitational force for each node $j$ and each mobile mass unit $\boldsymbol{x}_i(t)$  becomes
\begin{equation*} \label{eq:totalForceDistr}
\boldsymbol{f}_{\text{grav},i,j}(t)=\sum_{l\in \mathcal{B}_j} \sum_{n=1}^t g \cdot m_{i}(t) \cdot m_d \cdot  \frac{\boldsymbol{d}_l(n)-\boldsymbol{x}_i(t)}{\| \boldsymbol{d}_l(n)-\boldsymbol{x}_i(t)\|^p}.
\end{equation*}
In order to save communication cost, it is also possible to exchange aggregated information, e.g., the current estimates of the cluster centroids $\hat{\mathbf{w}}_k(t)$, instead of the feature vectors. If the exchange of $\mathbf{d}_{j}(t), j\in \mathcal{B}_j$ is left out, $\mathbf{D}_{j}(t)$ contains only the feature vectors observed by node $j$ itself, i.e.,  $\mathbf{D}_{j}(t)=\mathbf{d}_{j}(t)$, and the method proceeds analogously. \\

The intermediate cluster number estimate $\hat{K}_j^0(t)$ is improved upon by including received neighboring estimates $\hat{K}_{l}^0(t), {l}\in \mathcal{B}_j$ which are combined to form the final decision. To robustify against false decisions at the node level within the network, the ``combine'' step at node $j$ is chosen as 

\begin{equation} \label{eq:med}
\hat{K}_j(t)=\mathrm{median}\{\hat{K}_{l}^0(t)\},\quad l\in\mathcal{B}_j.
\end{equation}
An overview of other possible combination rules for the ``combine'' step of the diffusion algorithm is given, e.g., in \cite{sayed2014adaptation}.

As data streams in, the steps shown in Fig.\ref{Fig:ATCDiff} are repeated to provide an online in-network estimate $\hat{K}_j(t)$.\\

The D-GC algorithm is summarized in Algorithm \ref{DGC}.

  \begin{algorithm}
\caption{Distributed Gravitational Clustering (D-GC)} \label{DGC}
\begin{algorithmic}

\For {$t=1,..,N$}
     \ForAll {$j=1,...,J$}
          \State save collected feature vectors in $\mathbf{D}_j(t)$
      \EndFor
      
      \ForAll {$j=1,...,J$}
          \State exchange $\mathbf{d}_{j}(t)$ within $\mathcal{B}_j$ and update $\mathbf{D}_j(t)$
      \EndFor
      
      \ForAll {$j=1,...,J$}
          \State run GC as described in Algorithm \ref{GC1} 
         \State determine $\hat{K}^0_j(t)$ via Eq. (\ref{eq:massthresh})
      \EndFor
      
      \ForAll {$j=1,...,J$}
          \State exchange $\hat{K}_j^0(t)$ within $\mathcal{B}_j$
      \EndFor
      
      \ForAll {$j=1,...,J$}
          \State determine $\hat{K}_j(t)$ via Eq. (\ref{eq:med})
          \State estimate cluster centroids via Eq. (\ref{eq:estcent}) 
      \EndFor

\EndFor

\end{algorithmic}
\end{algorithm}

\section{Numerical Experiments} \label{sec:results}

This section evaluates the performance of the proposed method by using Monte Carlo experiments. It is benchmarked against the X-Means \cite{XMeans} and PG-Means \cite{PGMeans} cluster enumeration algorithms. Both methods are extensions of the K-Means algorithm which estimate the number of clusters based on applying model selection in a specified range $K_{\text{min}}<K<K_{\text{max}}$. To select a model amongst the candidates, X-Means applies the Bayesian information criterion (BIC), while PG means uses statistical hypothesis tests on one-dimensional projections of the data
and the model to determine if the examples are well represented by the model. For the distributed network setting, the proposed approach is benchmarked against the distributed extensions of \cite{XMeans,PGMeans}, i.e., the DX-Means and the DPG-Means that were recently proposed in \cite{Tekle_distrXMeans}.

\subsection{Simulation Setup} \label{sec:setup}
The data sets used for our simulations are a two-dimensional data set (Data-1) consisting of $K=5$ clusters as well as a more challenging data set with feature vectors of dimension $q=3$ and $K=6$ clusters (Data-2) with cluster centroids and covariance matrices as given in the Appendix. \\

The simulations are performed with the following parameters for GC: one mass unit with position $\mathbf{x}_i$ is placed around each new feature vector whereas its position is drawn at random from a Gaussian distribution with $\mathcal{N}(\mathbf{d}_n,r_x\mathbf{I}_q)$ and $r_x=1$. The parameters of GC are set as follows: $g$ is set to $1$, $k_{\text{damp}}=0.8$, $\epsilon_r= \sqrt{q}$ with $q$ being the data dimension, and $m_{\text{min}}=7$. 
The parameter $$p \equiv p(\mathbf{x}_1,\mathbf{x}_2)=(\log_{10} \big( \lVert \mathbf{x}_1-\mathbf{x}_2\rVert +1 \big)+2$$ is chosen throughout the simulations which is close to $p=2$ for close distances, but weakens the influence of far away clusters.\\
The minimum and maximum number of clusters which is considered for model order selection of X-Means and PG-Means are set to $K_{\text{min}}=1$ and $K_{\text{max}}=9$, respectively. 

The generation of outliers considers a certain percentage of samples that are affected by a process which is modeled by an additive contaminative distribution \cite{RobEst} based on Eq.~(\ref{eq:regmod}) with
\begin{equation}
\label{eq:outmod}
\boldsymbol{n}_{kj}(t)= \boldsymbol{e}_{kj}(t) +\zeta(t)\boldsymbol{o}_{kj}(t),
\end{equation}
where $\boldsymbol{e}_{kj}(t)$ represents an uncertainty term which models uncertainties following a specific probability distribution $E$, e.g. the Gaussian, while $\boldsymbol{o}_{kj}(t)$ denotes the contaminating outlier process that is independent of $\boldsymbol{e}_{kj}(t)$ and $\zeta(t)$ is a stationary random process for which 
\begin{equation*} 
\zeta(t)=
\begin{cases}
1 \quad \text{with probability } p_e\\
0 \quad \text{with probability } (1-p_e)
\end{cases}.
\end{equation*}
The displayed results represent the averages that are based on $100$ Monte-Carlo runs.

\subsection{Robustness Against Outliers}\label{sec:SimRob}
In order to evaluate the performance of GC in the presence of outliers or when the feature vectors do not follow a Gaussian distribution, the behavior of the algorithms for different underlying probability distributions is tested for the two data sets as a function of the number of available feature vectors per cluster. Since the benchmark algorithms are not adaptive, we let the algorithms under test process data batches in steps of $10$ at once for a better comparability, i.e., they are performed based on $10$, $20$, $30,...,K\cdot 50$ available feature vectors per cluster. Starting with one initial cluster, after every $50$ feature vectors per cluster, the number of clusters is increased by one and feature vectors belonging to the new clusters stream in. This ground truth for $K(t)$ is visualized by the solid black line, e.g. in Figs.~\ref{Fig:Data1} and \ref{Fig:Data2}.

\subsubsection{Data-1}
We first assume that no outliers are present and that the feature vectors are Gaussian distributed with $\mathbf{d}_n \sim \mathcal{N}(\boldsymbol{w}_k,\boldsymbol{\Sigma}_k)$, where $\boldsymbol{w}_k$ and $\boldsymbol{\Sigma}_k$ are given in the Appendix. The result is depicted in Figure \ref{Fig:Data1} (a), where the error bars visualize the estimation error and have a distance of each one standard deviation above and below the curve.
Except for some difficulties of the X-Means for a single cluster, the algorithms show similar and satisfactory results. \\
However, in real-life scenarios, the assumption of Gaussian-distributed data does not always hold \cite{RobEst}. The results for Laplace-distributed feature vectors while retaining the same parameters as before is presented in Fig.~\ref{Fig:Data1} (b). The figure shows the high sensitivity of the PG-Means to deviations from Gaussianity.
\begin{figure}
\centering
\begin{subfigure}[b]{.73\linewidth}
 \includegraphics[width=\textwidth]{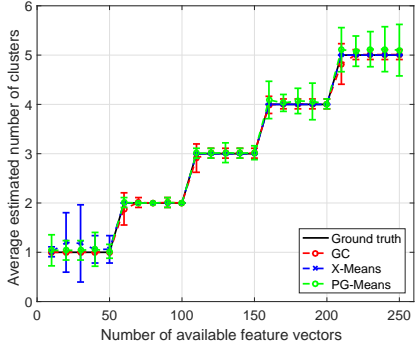}
   \caption{Average estimated number of clusters as a function of the number of feature vectors.}
\end{subfigure}
\hfill
\begin{subfigure}[b]{.73\linewidth}
  \includegraphics[width=\textwidth]{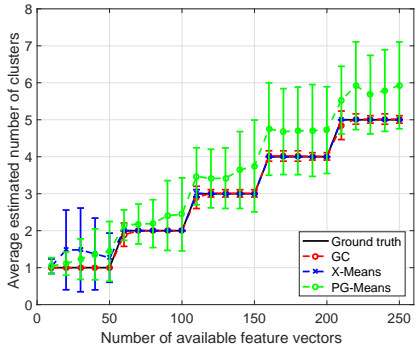}
  \caption{Average estimated number of clusters as a function of the number of feature vectors, where the feature vectors follow a Laplace distribution.}
\end{subfigure}

\begin{subfigure}[b]{.73\linewidth}
 \includegraphics[width=\textwidth]{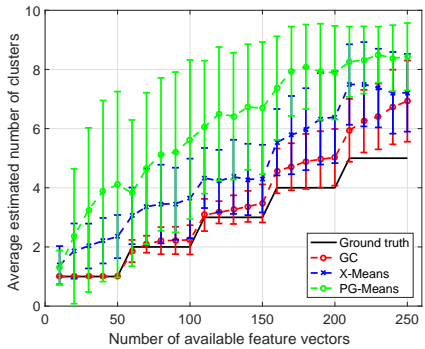}
  \caption{Average estimated number of clusters as a function of the number of feature vectors, where the feature vectors are contaminated with $5 \%$ chi-square distributed additive outliers.}
\end{subfigure}

\caption{Evaluation of Data-1 for different underlying probability distributions for the feature vectors and outliers.}
\label{Fig:Data1}

\end{figure}

Since real measurements may also suffer from outlying data, the following simulation evaluates the performance of the algorithms under test for the presence of $p_e=5 \%$ outliers in the feature vectors. The outliers are generated by drawing a sample at random from a probability distribution and adding it to the original sample (additive outliers) as specified in Eq. (\ref{eq:outmod}). %
In real-world scenarios, outliers usually do not follow any specific distribution. We choose $\boldsymbol{o}_{kj}(t)$ to follow a non-symmetric distribution, i.e., a skewed heavy-tailed distribution like the chi-square distribution with different degrees of freedom $v$ for each cluster. This is done in order to create a non-symmetric outlier distribution instead of a constant shift of the mean of the outlier distribution for all clusters. In this manner, a randomly drawn vector of dimension $q$ with $v_1=3$ is added to 5 $\%$ of data vectors for the first cluster and a vector with $v_2=5$ is subtracted from the corresponding feature vectors of cluster~$2$. For cluster~$3$, a different random number is drawn for each direction in feature space, i.e. the outliers are generated with $v_{3,1}=4$ and $v_{3,2}=1$ for $x$ and $y$ direction, respectively, whereby $v_{3,2}=1$ is subtracted from the $y$-component. For the fourth cluster, we have $v_{4,1}=2$, $v_{4,2}=-3$ and the outliers for the last cluster are generated by drawing samples at random from a Gaussian distribution such that $\boldsymbol{o}_{kj} \sim \mathcal{N}(\boldsymbol{w}_o, \Sigma_o)$ with $\boldsymbol{w}_o=(0, 0)^\top$ and $\Sigma_o=3 \mathbf{I}_2$. The result is depicted in Fig.~\ref{Fig:Data1} (c). A small amount of outliers in the data causes significantly worse estimation results for both the X-Means and PG-Means. The increase in the estimation error is less pronounced for the proposed GC algorithm.

\subsubsection{Data-2}
\begin{figure}%
\centering
\begin{subfigure}[b]{.75\linewidth}
 \includegraphics[width=\textwidth]{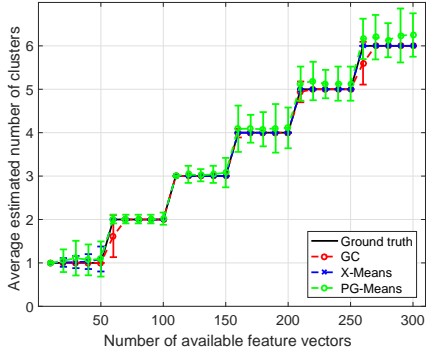}
   \caption{Average estimated number of clusters as a function of the number of feature vectors.}
\end{subfigure}
\hfill
\begin{subfigure}[b]{.75\linewidth}
  \includegraphics[width=\textwidth]{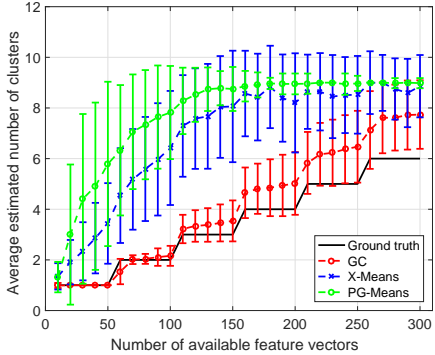}
  \caption{Average estimated number of clusters as a function of the number of feature vectors per node, where the feature vectors are contaminated with $5 \%$ Gaussian-distributed additive outliers.}
\end{subfigure}
\caption{Evaluation of Data-2 with and without outliers.}
\label{Fig:Data2}
\end{figure}

For the second, more complex data set with $\boldsymbol{w}_k$ and $\boldsymbol{\Sigma}_k$ as given in the Appendix, we evaluate the performance of GC, X-Means and PG-Means analogously as for Data-1 for the outlier-free case and for the presence of $p_e=5 \%$ Gaussian-distributed additive outliers. 
The result for outlier-free feature vectors is depicted in Fig.~\ref{Fig:Data2}~ (a) and Fig.~\ref{Fig:Data2}~(b) presents the estimation outcome for $5 \%$ Gaussian-distributed outliers with $\boldsymbol{o}_{kj} \sim\mathcal{N}(\boldsymbol{w}_o, \Sigma_o)$ and $\boldsymbol{w}_o=(0, 0,0)^\top$ and $\Sigma_o=3 \mathbf{I}_3$.

The corresponding RMSE and probability of correct detection results are given in Tables II and III.
 \begin{table}\label{tab:RMSE}
\centering
\normalsize
\begin{tabular}{lccc} 
\toprule
Simulation &\multicolumn{3}{c}{RMSE} \\ 
&GC & X-Means & PG-Means \\ 
\midrule 
Fig.~\ref{Fig:Data1} (a) &0.152 & 0.220 &0.267  \\
Fig.~\ref{Fig:Data1} (b) &0.150 & 0.425 &1.092  \\
Fig.~\ref{Fig:Data1} (c) &1.065 & 2.077 &3.841  \\
\midrule
Fig.~\ref{Fig:Data2} (a) &0.168 & 0.071 &0.376  \\
Fig.~\ref{Fig:Data2} (b) &1.317 & 3.986 &4.711  \\
\bottomrule
\end{tabular}
\caption{RMSE of the estimated number of cluster centroids compared to the ground truth for the algorithms under test for different simulation scenarios.}
\end{table}

\begin{table}\label{tab:probCorrDec}
\centering
\normalsize
\begin{tabular}{lccc} 
\toprule
Simulation &\multicolumn{3}{c}{$p_\text{corr}$} \\ 
&GC & X-Means & PG-Means \\ 
\midrule 
Fig.~\ref{Fig:Data1} (a) &0.977 & 0.986 &0.9672  \\
Fig.~\ref{Fig:Data1} (b) &0.978 & 0.964 &0.704  \\
Fig.~\ref{Fig:Data1} (c) &0.647 & 0.207 &0.140  \\
\midrule
Fig.~\ref{Fig:Data2} (a) &0.972 & 0.995 &0.928  \\
Fig.~\ref{Fig:Data2} (b) &0.605 & 0.088 &0.054  \\

\bottomrule
\end{tabular}
\caption{Probability of correct decision $p_\text{corr}$ for the algorithms under test for different simulation scenarios.}
\end{table}

\subsection{Distributed Cluster Enumeration} 
In the following scenario, the performance of the proposed algorithm is tested for a distributed WSN where $J=10$ nodes are randomly distributed in space. Each node is connected to the $|\mathcal{B}_j|=4$ neighboring nodes which have the smallest Euclidean distance. For the clustering procedure, again Data-2 is used. Each node is subject to a contamination with a certain amount of additive Gaussian-distributed outliers following a Gaussian distribution with $\mathbf{w}_o=(0, 0,0)^\top$ and $\Sigma_o= 3 \mathbf{I}_3$: three nodes receive outlier-free feature vectors, three nodes receive feature vectors including $5\%$ outliers, three nodes are subject to $10\%$ outliers and one node suffers from $20\%$ outlier contamination. 

\begin{figure}%
\centering
\begin{subfigure}[b]{.73\linewidth}
 \includegraphics[width=\textwidth]{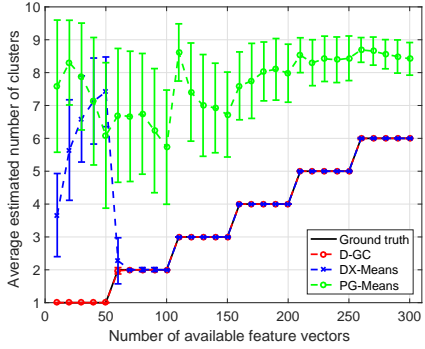}
   \caption{Average estimated number of clusters as a function of the number of feature vectors including both exchange steps.}
\end{subfigure}
\hfill
\begin{subfigure}[b]{.73\linewidth}
  \includegraphics[width=\textwidth]{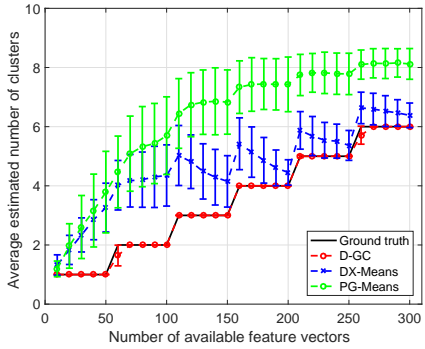}
  \caption{Average estimated number of clusters as a function of the number of feature vectors including the exchange of the number of estimated clusters $\hat{K}_{j}(t)$ only.}
\end{subfigure}

\begin{subfigure}[b]{.73\linewidth}
 \includegraphics[width=\textwidth]{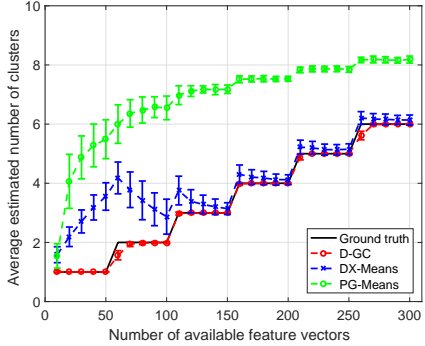}
  \caption{Average estimated number of clusters as a function of the number of feature vectors for a non-cooperative network.}
\end{subfigure}

\caption{Evaluation of Data-2 for a distributed network compared for different variations of data exchange: the exchange of both feature vectors and estimates of the number of clusters (a), exchanging the estimated number of clusters only (b) and a non-cooperative scenario with no data exchange at all (c).}
\label{Fig:DistrScen}

\end{figure}

Fig.~\ref{Fig:DistrScen} shows the cluster enumeration results for three different amounts of data exchange in the diffusion adaptation algorithm of the distributed clustering algorithms.  Fig.~\ref{Fig:DistrScen} (a) shows the results when the feature vectors and the cluster number estimates $\hat{K}_j(t)$ are communicated within each neighborhood $\mathcal{B}_j$. Fig.~\ref{Fig:DistrScen} (b) displays the results, when only the estimates $\hat{K}_j(t)$ are exchanged and Figure~\ref{Fig:DistrScen} (c) considers the case of a non-cooperative network, i.e., the nodes do not communicate at all. It can be seen that the propagation of outlier-contaminated data can result in worse clustering compared to the non-cooperative network scenario. In contrast to its competitors, GC is able to correctly estimate the number of clusters for all variations of data exchange in the evaluated scenario. 

As GC provides not only an estimate of the number of clusters but also of the location of the cluster centroids, the following simulation evaluates the accuracy of the location estimates. As a benchmark, this paper considers the \textit{Distributed K-Means} (DKM) algorithm by Forero {\it et al.} \cite{DKM}. The basic idea of the DKM is to cluster the features into a preset number of groups, such that the sum of squared-errors is minimized. For details, see \cite{DKM}. Table III provides the root-mean-square error (RMSE) of the estimated cluster centroids for both GC and DKM compared to the ground truth for this scenario. The results for the DKM are based on 20 and 50 iterations with parameters $p=\nu=2$, where $p=2$ enables soft clustering and $\nu=2$ is the tuning parameter which yields the best results in the performance tests in \cite{DKM}. As a result, GC is able to provide precise estimates of the cluster centroid locations. In the given scenario, it outperforms the benchmark method with a RMSE being constantly low, independently of the number of available feature vectors per cluster.

The RMSE and probability of correct detection for the distributed clustering scenario is presented in Tables~IV and V, respectively.

\begin{table}\label{tab:clusterCentroids}
\centering
\normalsize
\begin{tabular}{lcccc} 
\toprule
&\multicolumn{4}{c}{Number of feature vectors per cluster}\\ 
\midrule 
&10 & 20 & 50&100\\ 
\midrule 
D-GC &0.1814 & 0.1754 &0.1851 & 0.1752 \\
DKM 20 it.& 2.5865&1.9133&1.4923&1.7282\\
DKM 50 it.& 1.3925&0.6847&0.1811&0.3525\\
\bottomrule
\end{tabular}
\caption{RMSE of the estimated cluster centroids compared to the ground truth for  different amounts of feature vectors per cluster.}
\end{table}

\begin{table}\label{tab:RMSE_distr}
\centering
\normalsize
\begin{tabular}{lccc} 
\toprule
Simulation &\multicolumn{3}{c}{RMSE} \\ 
&GC & X-Means & PG-Means \\ 
\midrule 
Fig.~\ref{Fig:DistrScen} (a) &0.035 & 2.269 &4.736  \\
Fig.~\ref{Fig:DistrScen} (b) &0.147 & 1.839 &3.649  \\
Fig.~\ref{Fig:DistrScen} (c) &0.201 & 1.682 &4.335  \\
\bottomrule
\end{tabular}
\caption{RMSE of the estimated number of cluster centroids compared to the ground truth for the algorithms under test for the distributed clustering scenario.}
\end{table}

\begin{table}\label{tab:probCorrDec_distr}
\centering
\normalsize
\begin{tabular}{lccc} 
\toprule
Simulation &\multicolumn{3}{c}{$p_\text{corr}$} \\ 
&GC & X-Means & PG-Means \\ 
\midrule 
Fig.~\ref{Fig:DistrScen} (a) &0.999 & 0.832 &0.002  \\
Fig.~\ref{Fig:DistrScen} (b) &0.978 & 0.439 &0.283  \\
Fig.~\ref{Fig:DistrScen} (c) &0.960 & 0.772 &0.303  \\
\bottomrule
\end{tabular}
\caption{Probability of correct decision $p_\text{corr}$ for the algorithms under test for the distributed clustering scenario.}
\end{table}

%
%

\subsection{Convergence}
In order to demonstrate the convergence of GC after a limited number of iterations, the temporal behavior of the mass units, which are placed in different initial distances to the feature vectors, is evaluated. For this purpose, a single cluster with 50 feature vectors, $\mathbf{d}_n \sim \mathcal{N}(\boldsymbol{w}_k,\boldsymbol{\Sigma}_k)$, is generated with $\boldsymbol{w}=(3, 3)^\top$ and covariance matrix $ \boldsymbol{\Sigma}= (0.3, 0.3)^\top \mathbf{I}_2$. One mobile mass unit per feature vector is placed in feature space with an associated position that is drawn at random from a Gaussian distribution with $\mathcal{N}(\mathbf{d}_n,\sigma\mathbf{I}_2)$. GC is executed and the average Euclidean distance between the mobile mass units and the cluster centroid is calculated for each time step. As the mobile mass units approach the cluster, their distance to the centroid decreases continuously until a minimum distance $\epsilon_{\text{min}}$ is reached, after which the algorithm is said to have converged. 

\begin{figure}
    \centering
  \includegraphics[width=.4\textwidth]{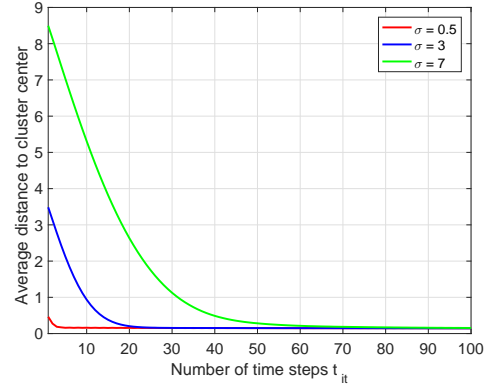}
    \caption{Average distance of mass units to cluster center as a function of the number of time steps $t_{\text{it}}$.}
    \label{Fig:limDist2cluster}
\end{figure}

Fig.~\ref{Fig:limDist2cluster} provides the average number of time steps $t_{\text{it}}$ that the GC algorithm requires to achieve convergence for the cases $\sigma=0.5$, $\sigma=3$ and $\sigma=7$. 
It is shown that the mobile mass units converge to the true cluster centroid during the clustering process and remain in their position even for large initial distances from the cluster centroid. Small values of $\sigma$ result in lower computation times.

%
%

\subsection{Computational Cost} \label{Sec:compTime}
\begin{figure}
    \centering
  \includegraphics[width=.4\textwidth]{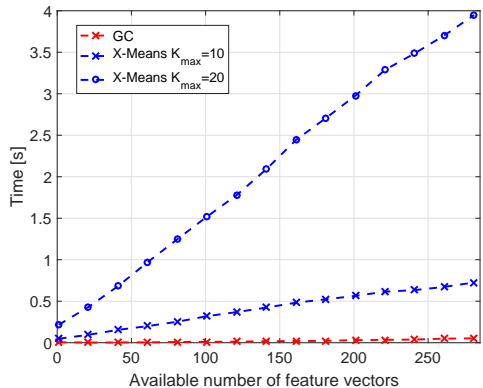}
    \caption{Comparison of computation times for GC and X-Means for different values of $K_{\text{max}}$.}
    \label{Fig:compT}
\end{figure}

Since the energy consumption in wireless devices should be kept as low as possible, the computational cost of the algorithms in use is crucial. In order to compare the algorithms in this aspect, the computation time as a function of the number of available feature vectors is provided by Fig.~\ref{Fig:compT} and given in seconds (using an Intel Core i5 Quad Core 760 PC1156). Because of its performance results and disproportionately high running times, the PG-Means is not considered in this simulation.\\

\section{Multi-View Camera Network Application}\label{sec:camNet}

\begin{figure}
    \centering
  \includegraphics[width=.4\textwidth]{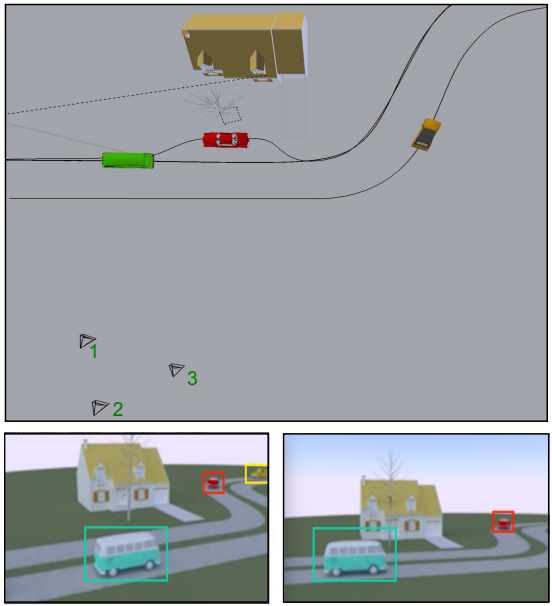}
    \caption{A wireless cooperative camera network continuously observing a scene. The top image depicts the setup with $J=3$ cameras observing the scene from different viewpoints. The bottom left and right images show the frames captured at the same time instant by camera 1 and 3, respectively. }
    \label{Fig:camNet}
\end{figure}

The performance of the D-GC algorithm in a cooperative WSN application is evaluated based on the following multi-camera video example introduced in \cite{Tekle_distrXMeans}. We consider a wireless camera network as shown in Fig.~\ref{Fig:camNet} where a set of spatially distributed cameras (nodes) records the scene from different angles. By communicating with each other, the aim is to adaptively determine a network-wide estimate of the number of recorded cars.
Due to the different viewpoints, the number of observed objects per camera at the same time instant can differ significantly. For this reason, we selected cameras 1-3 to cooperate such that $J=| \mathcal{B}_j | =3$: camera 1 and 2 have a similar view on the scene whereas camera 3 does not see all of the cars recorded by camera 1 and 2. Plus, the cars enter and leave the scene with a delay compared to the other cameras.\\  A video of $95$ frames is recorded by the cameras where three cars enter the scene at different time instants. A Gaussian Mixture Model foreground detector is used to separate moving objects from the background in order to extract feature vectors for the clustering process. The feature vectors are composed of SURF \cite{SURF} and color features. For the color histogram, the detected object is subdivided into three concentric rings. For each ring, a 10-bin histogram per color channel is computed. Concatenating the three histograms for each of the three color channels results in a 90-dimensional color feature. Combining it with the SURF features yields a 211-dimensional feature vector for each object. It is a challenging scenario since the video has a low resolution, the cameras observe the cars from different angles and there is a low amount of available feature vectors.
Because of the previous results, we will compare only DX-Means and D-GC in the following. For this scenario, the feature vectors lie close in space which is taken into account by setting the parameters to $K_{\text{max}}=5$, $r_x=0.01$, $m_{\text{min}}=2$, $d_{\text{max}}=3$ and $\epsilon_t=0.12$. The cameras exchange their estimates of the number of object (cars) in the scene and decide on a common estimate according to Eq. (\ref{eq:med}). \\
\begin{figure}
    \centering
  \includegraphics[width=.4\textwidth]{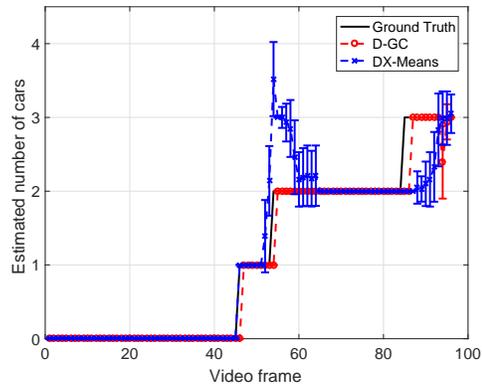}
    \caption{Cluster enumeration result for the cooperative camera network.}
    \label{Fig:ResCamNet}
\end{figure}
The average results based on $100$ Monte-Carlo runs are provided in Fig.~\ref{Fig:ResCamNet}. The variance is a result of the random initialization of the DX-Means and the random initial positioning of mobile mass units for GC. While the DX-Means needs time to converge to the true value, the D-GC provides a close representation of the ground truth as seen by camera~2.

%
%
\section{Conclusion} 
A novel clustering algorithm was proposed, which can be adapted to various single-node and in-network scenarios. It provides both an estimate for the number of clusters as well as for the positions of their centroids. A performance comparison to the X-Means and PG-Means has been provided where our proposed method shows promising results. Unlike the benchmarks, the GC is potentially real-time capable and, up to a certain amount, robust against outliers in the data. Using a standard set of default parameters provides good results for the diverse scenarios examined in this work. By including knowledge about the data set in defining the parameters, GC can be further adapted to very specific scenarios, resulting in an increased performance.

Future work will include the application of this algorithm to real-world speech source labelling for distributed signal enhancement multi-device-multi-task (MDMT) wireless acoustic sensor networks \cite{bertrand2011distributed}, object labelling in distributed multi-view camera networks  as well as labelling of semantic information based on occupancy grid maps for autonomous mapping and navigation with multiple rescue robots \cite{Kohlbrecher2014hector}.

\section*{Acknowledgments}
The work of P. Binder was supported by the LOEWE initiative (Hessen, Germany)  within  the  NICER  project. The work of M. Muma was supported by the project HANDiCAMS which acknowledges the financial support of the Future and Emerging Technologies (FET) programme within the Seventh Framework Programme for Research of the European Commission, under FET-Open grant number: 323944.

\appendix \label{app}

The cluster centroids and diagonal entries of the corresponding covariance matrices for the data sets Data-1 and Data-2 as introduced in Subsection \ref{sec:setup} are given as follows: \\
\\
\textit{Data-1}:\\
$\boldsymbol{w}_1=(-1, 0)^\top$, $\boldsymbol{w}_2=(4, 0)^\top$, $\boldsymbol{w}_3=(0,5)^\top$, $\boldsymbol{w}_4=(9,4)^\top$, $\boldsymbol{w}_5=(3, 9)^\top$,\\
 $ \boldsymbol{\Sigma}_1=(0.2, 0.4)^\top \mathbf{I}_2$, $ \boldsymbol{\Sigma}_2=(0.6, 0.6)^\top \mathbf{I}_2$, $ \boldsymbol{\Sigma}_3=(0.4, 0.2)^\top \mathbf{I}_2$, $ \boldsymbol{\Sigma}_4=(0.2, 0.2)^\top \mathbf{I}_2$, $ \boldsymbol{\Sigma}_5=(0.3, 0.5)^\top \mathbf{I}_2$.
\\
\\
\textit{Data-2}: \\ $\boldsymbol{w}_1=(-1, 0, 7)^\top$, $\boldsymbol{w}_2=(3, 0, 8)^\top$, $\boldsymbol{w}_3=(0,5,1)^\top$, $\boldsymbol{w}_4=(9,4,4)^\top$, $\boldsymbol{w}_5=(3, 9,5)^\top$, $\boldsymbol{w}_6=(5, 5,1.55)^\top$, \\ $ \boldsymbol{\Sigma}_1=\alpha (0.2, 0.4, 0.2)^\top \mathbf{I}_3$, $ \boldsymbol{\Sigma}_2=\alpha (0.6, 0.3,0.5)^\top \mathbf{I}_3$, $ \boldsymbol{\Sigma}_3=\alpha (0.4, 0.2,0.1)^\top \mathbf{I}_3$, $ \boldsymbol{\Sigma}_4=\alpha (0.3, 0.3, 0.3)^\top \mathbf{I}_3$, $ \boldsymbol{\Sigma}_5=\alpha (0.3, 0.5, 0.3)^\top \mathbf{I}_3$, $ \boldsymbol{\Sigma}_6=\alpha (0.4, 0.4, 0.4)^\top \mathbf{I}_3$ with $\alpha=0.15$.

\bibliographystyle{IEEEtran}
\bibliography{Literatur}

\end{document}